\documentclass[aps,pra,twocolumn,superscriptaddress,epsfig,groupaddress]{revtex4-1}

\usepackage{amsfonts}
\usepackage{hyperref}
\usepackage{amsmath}
\usepackage{amssymb}
\usepackage{graphicx}
\usepackage{epsfig}
\usepackage{color}
\usepackage{dsfont}
\usepackage{soul}
\usepackage{amstext,amsthm,graphics,dcolumn,bm,latexsym,hyperref,subfigure}

\newcommand\tr{{\rm Tr}\;}

\begin{document}

\title{Quantum uncertainty in critical systems with three spins interaction}

\author{Thiago M. Carrijo}
\affiliation{Instituto de F\'{i}sica, Universidade Federal de Goi\'{a}s, 74.001-970, Goi\^{a}nia, Goi\'{a}s, Brazil}

\author{Ardiley T. Avelar}
\affiliation{Instituto de F\'{i}sica, Universidade Federal de Goi\'{a}s, 74.001-970, Goi\^{a}nia, Goi\'{a}s, Brazil}

\author{Lucas C. C\'{e}leri}
\email{lucas@chibebe.org}
\affiliation{Instituto de F\'{i}sica, Universidade Federal de Goi\'{a}s, 74.001-970, Goi\^{a}nia, Goi\'{a}s, Brazil}

\begin{abstract}
In this article we consider two spin$-1/2$ chains described, respectively, by the thermodynamic limit of the $XY$ model with the usual two site interaction, and an extension of this model (without taking the thermodynamics limit), called $XYT$, were a three site interaction term is presented. To investigate the critical behaviour of such systems we employ tools from quantum information theory. Specifically, we show that the local quantum uncertainty, a quantity introduced in order to quantify the minimum quantum share of the variance of a local measurement, can be used to indicate quantum phase transitions presented by these models at zero temperature. Due to the connection of this quantity with the quantum Fisher information, the results presented here may be relevant for quantum metrology and quantum thermodynamics.
\end{abstract}

\maketitle

\section{Introduction}

Characterized by a change in the macroscopic order of a physical system, phase transitions are ubiquitous in nature. Contrary to thermodynamical phase transitions, which occur at finite temperature, quantum phase transitions (QPTs) take place at zero temperature, driven by quantum fluctuations \cite{sachdev}. Indeed, being linked with energies level crossings in the ground state, QPTs can occur even at finite temperature, as long as the system cannot be excited by the thermal fluctuations. A candidate system to study these kinds of phenomena is a quantum spin chain, which can present several distinct phases \cite{sachdev}.

Many interesting physical systems exhibit QPTs, among which we mention the superfluid to insulating \cite{fisher2} and the superconducting to insulating \cite{liu2} transitions for bosonic systems while, for fermionic ones, we have the spin wave density to a magnetic phase \cite{mathur,Lohneysen07} and the quantum spin glass to paramagnetic phase transitions \cite{sachdev2} as examples. A remarkable feature of QPTs is that, in the neighbourhood of a critical point, the system becomes strongly correlated, and its correlation length diverges at this point \cite{sachdev}. This has suggested that quantum correlations should be maximum near the transition point, fact that was verified in seminal works about entanglement measures in critical systems \cite{osterloh,osborne} (see Ref. \cite{amicoRev} for a review regarding entanglement in many-body systems). As a natural consequence, other measures of quantum correlations were also employed to investigate the properties of QPTs, such as quantum discord \cite{sarandy1,maziero,maziero1,sarandy3,werlang,werlang1} and related quantities \cite{sarandy2,justino,cheng,shan}.

Within the last decade, quantum correlations other than entanglement were discovered \cite{ollivier,vedral} and their importance for quantum information processing has only increased \cite{mile,philip,datta08,lanyon08}. The origin of such correlations lies in the existence of quantum coherence among different partitions of a quantum system, being present even in separable states (see Ref. \cite{modi} for a recent review). Among all the quantum correlation measures, the so-called local quantum uncertainty (LQU) stands out as a bona fide measure with a good computability and a close mathematical expression for any qubit-qudit bipartite system \cite{girolami}. This measure is based on the skew information, introduced in Ref. \cite{wigner} to quantify the uncertainty in the measurement of an observable. The LQU is defined just as the minimazation of this quantity over all possible observables. Beyond its importance as a quantum correlation quantifier, LQU is also related with the quantum Fisher information \cite{luo,petz,luo2}, which makes it very important for quantum metrology. Moreover, it has been shown that it is also connected with the speed of evolution of a quantum system \cite{girolami}. These are some of the reasons that lead us to choose LQU as a tool to investigate critical behaviour. 

Here we study local quantum uncertainty in critical systems considering two spin-$1/2$ chains, described by the anisotropic $XY$ Hamiltonian and the $XY$ model with three spins interaction, named $XYT$ model \cite{suzuki,titvinidze,li}, which contains the spin-liquid phase \cite{anderson} that only recently could be experimentally investigated \cite{balents}. We consider the zero temperature case, making some comments about finite temperature in the last section.

Several works have addressed the critical behaviour of the $XY$ model from the point of view of quantum information, considering quantum and classical correlations quantifiers as tools for the identification of quantum phase transitions (see, e.g., Refs. \cite{sarandy3,modi} and references therein). Recently, the local quantum uncertainty was also computed for some instances of this model and its usefulness for the identification of the QPT was verified \cite{fanchini}. However, the case of the $XYT$ spin chain received much little attention. In Ref. \cite{li} the authors demonstrated that quantum discord \cite{ollivier} and classical correlations \cite{vedral} are good indicators for the critical behaviour of this model, while entanglement measures were considered in Ref. \cite{Zvyagin}. The purpose of the present article is to contribute to this debate by investigating how local quantum uncertainty behaves for the $XYT$ model. We also provide a complementary analysis of the usual $XY$ Hamiltonian, presented in Ref. \cite{fanchini}. While the $XY$ Hamiltonian is investigated in the thermodynamic system, for the $XYT$ model we consider a finite chain.

The paper is structured as follows. In Section II we review the definition of the local quantum uncertainty while Section III is devoted to present both models and our results. The final considerations are done in IV.

\section{Local Quantum Uncertainty}

The LQU was introduced in order to quantify the minimal quantum uncertainty that could be attainable in a measurement of a local observable \cite{girolami}. This minimal uncertainty comes from the existence of quantum correlations between the system being measured and the rest of the universe. Based on this fact, a measure of quantum correlations was then proposed \cite{girolami}. Let $\rho$ be the state of a bipartite system and $K$ some local observable on one of the partitions, from here on called the system. LQU is defined as
\begin{equation}
\mathcal{U}(\rho) \equiv \min_{K} \mathcal{I}(\rho, I\otimes K),
\label{LQU}
\end{equation}
where
\begin{equation*}
\mathcal{I}(\rho, I\otimes K)=-\dfrac{1}{2}\tr\left([\sqrt{\rho},K]^{2}\right) 
\end{equation*}
is the so-called skew information \cite{wigner,luo}, which is a measure of the non-commutativity between the state and the observable. The minimum is taken over the set of all observables acting on the considered partition. Strictly speaking, the definition of $\mathcal{U}$ should take into account the spectrum of the observable being measured. Each different choice of the spectrum leads us to a different family of measures. However, if the system is a qubit, as in the present article, the spectrum does not change the behaviour of $\mathcal{U}$ (except for a global multiplicative constant), therefore we do not explicitly considered it in the definition (\ref{LQU}), as originally done \cite{girolami}. Moreover, as we are dealing only with qubits (spin$-1/2$ particles), we consider the following general form for the local observable $K \equiv \vec{r}\cdot\vec{\sigma}$, where $\vec{\sigma}$ is the Pauli vector operator and $|\vec{r}|=1$. This enables one to compute a closed, analytical, form for LQU \cite{girolami}
\begin{equation*}
 \mathcal{U}(\rho) = 1 - \lambda_{max}\{W\},
\end{equation*}
where the matrix $W$ has the following elements
\begin{equation*}
 (W)_{ij} \equiv \tr\{\sqrt{\rho}(\sigma_{i}\otimes \mathds{I})\sqrt{\rho}(\sigma_{j}\otimes \mathds{I})\},
\end{equation*}
and $\lambda_{max}$ denotes the maximum eigenvalue of the matrix $W$. 

As LQU quantifies the minimum \emph{quantum uncertainty} in a measurement performed on the system, it is zero only when the state of the system commutes with some observable, which implies that the local system is not quantum correlated with anything else. Note that the existence of this uncertainty (even in the presence of a flawless apparatus) is caused only by the probabilistic character of the quantum measurement. In other words, if the system of interest is correlated with some other system, its reduced state cannot be an eigenstate of any local observable, thus inevitably leading to a non-vanishing variance.

In the following section we employ this tool to study the critical behaviour of two spin chains, showing that it is a very good candidate for pinpoint quantum phase transitions in both models.

\section{Model and results}

As stated in the Introduction, we consider two distinct models, the $XY$ spin chain with two and three spins interaction. The case of two spins interaction was recently considered in Ref. \cite{fanchini}. However, the focus of that work is the so called local quantum coherence, as measured by the skew information defined in the preceding section, in the $XY$ model. Although $\mathcal{U}$ is an extremization of $\mathcal{I}$, here we extend the analysis reported in Ref. \cite{fanchini} considering the entire range of the anisotropic parameter. Moreover, we focus in the case of three spins interaction model, which was not considered in Ref. \cite{fanchini}. 

\subsection{The $XY$ model with two spins interaction}

We describe here the paradigmatic example of the $XY$ model, which govern the dynamics of a spin$-1/2$ chain under the effect of a magnetic field in $z-$direction anisotropically interacting in the $xy$ plane. The Hamiltonian of this model is given by
\begin{eqnarray}
H &=& -\frac{1}{2}\sum_{j=1}^{N}\left[(1+\gamma)\sigma^{x}_{j}\sigma^{x}_{j+1}+(1-\gamma)\sigma^{y}_{j}\sigma^{y}_{j+1} + \lambda\sigma^{z}_{j} \right],
\label{HXY}
\end{eqnarray}
with $\gamma \in [0,1]$ and $N$ being, respectively, the anisotropy parameter and the dimension of the chain, and $\lambda$ stands for the strength of the external field. For $\gamma = 0$ the Hamiltonian reduces to the $XX$ model while for $\gamma=1$ we obtain the Ising Hamiltonian in a transverse field. For any other value ($0<\gamma<1$) $H$ belongs to the Ising universality class. $\lambda$ is the reciprocal of the external transverse magnetic field and $\sigma^{k}_{j}$ is the $i-$th Pauli matrix for spin $j$. 

Through Jordan-Wigner and Bogoliubov transformations, the Hamiltonian (\ref{HXY}) can be diagonalized, in the thermodynamic limit ($N\rightarrow \infty$), in terms of spinless fermions operators \cite{sachdev}. Remembering that the model posses translational and parity symmetries \cite{osborne}, the reduced density matrix of two arbitrary spins can be written as 
\begin{equation}
\rho_{0n} = \frac{1}{4}\left[I_{0n}+\langle\sigma^{z}\rangle(\sigma_{0}^{z}+\sigma_{n}^{z})+\sum_{k=1}^{3}\langle\sigma_{0}^{k}\sigma^{k}_{n}\rangle\sigma_{0}^{k}\sigma^{k}_{n}\right]. 
\label{densityXY}
\end{equation}
The details of the correlation functions appearing in the above equation are given in Appendix A. It is important to note here that we are dealing with the so-called thermal ground state, which is the zero temperature limit of the canonical ensemble. Moreover, we are disregarding the effects of symmetry breaking (see Ref. \cite{sarandy3} to further details regarding this issue).

This model presents a second order quantum phase transition at the critical point $\lambda_{c} = 1$, for all values of $\gamma$, which is of an order-disorder type, separating the ferromagnetic from the paramagnetic phases \cite{sachdev}.

\begin{figure}[h]
\begin{center}
\includegraphics[scale=0.37]{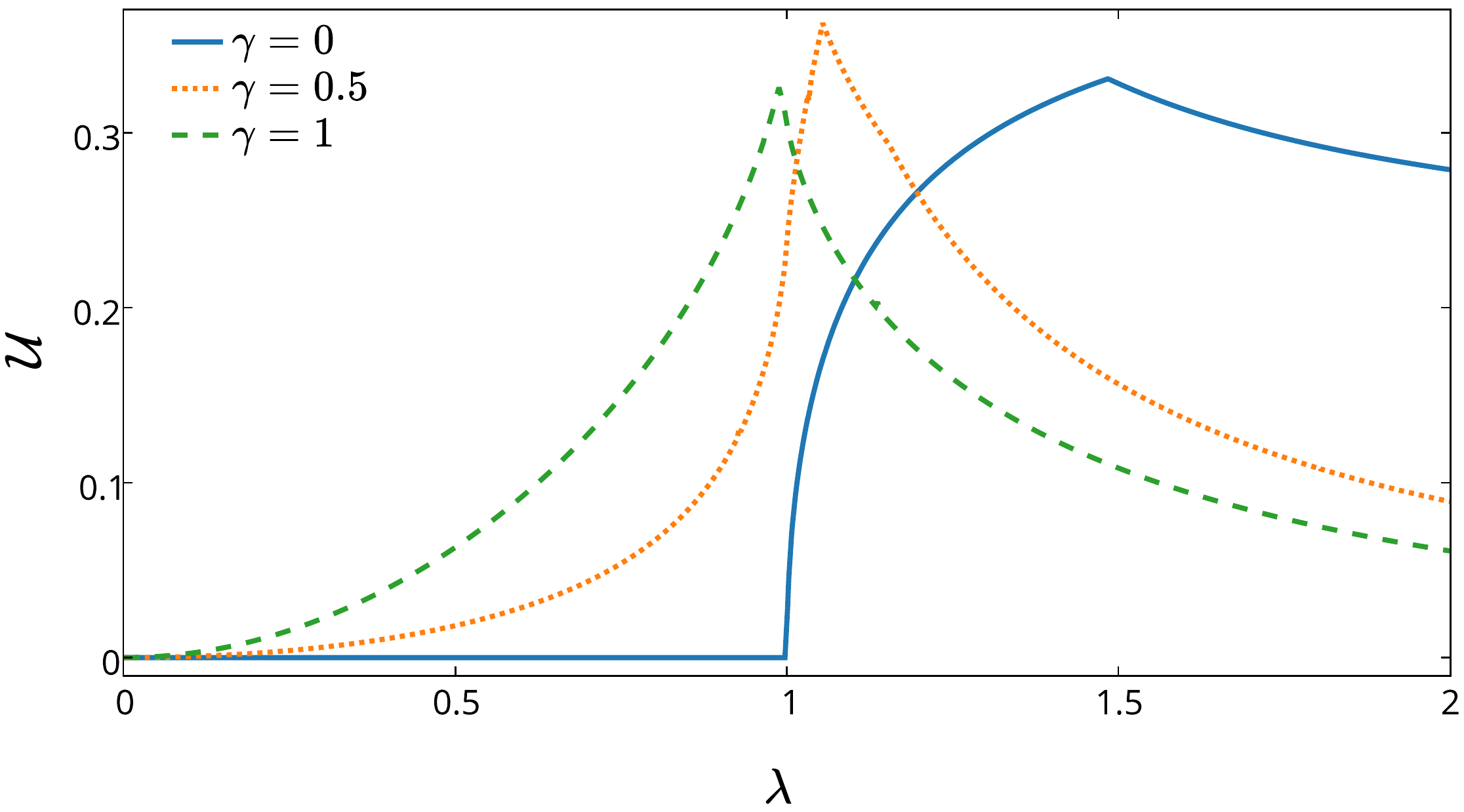} \\ \vspace{0.5cm}
\includegraphics[scale=0.37]{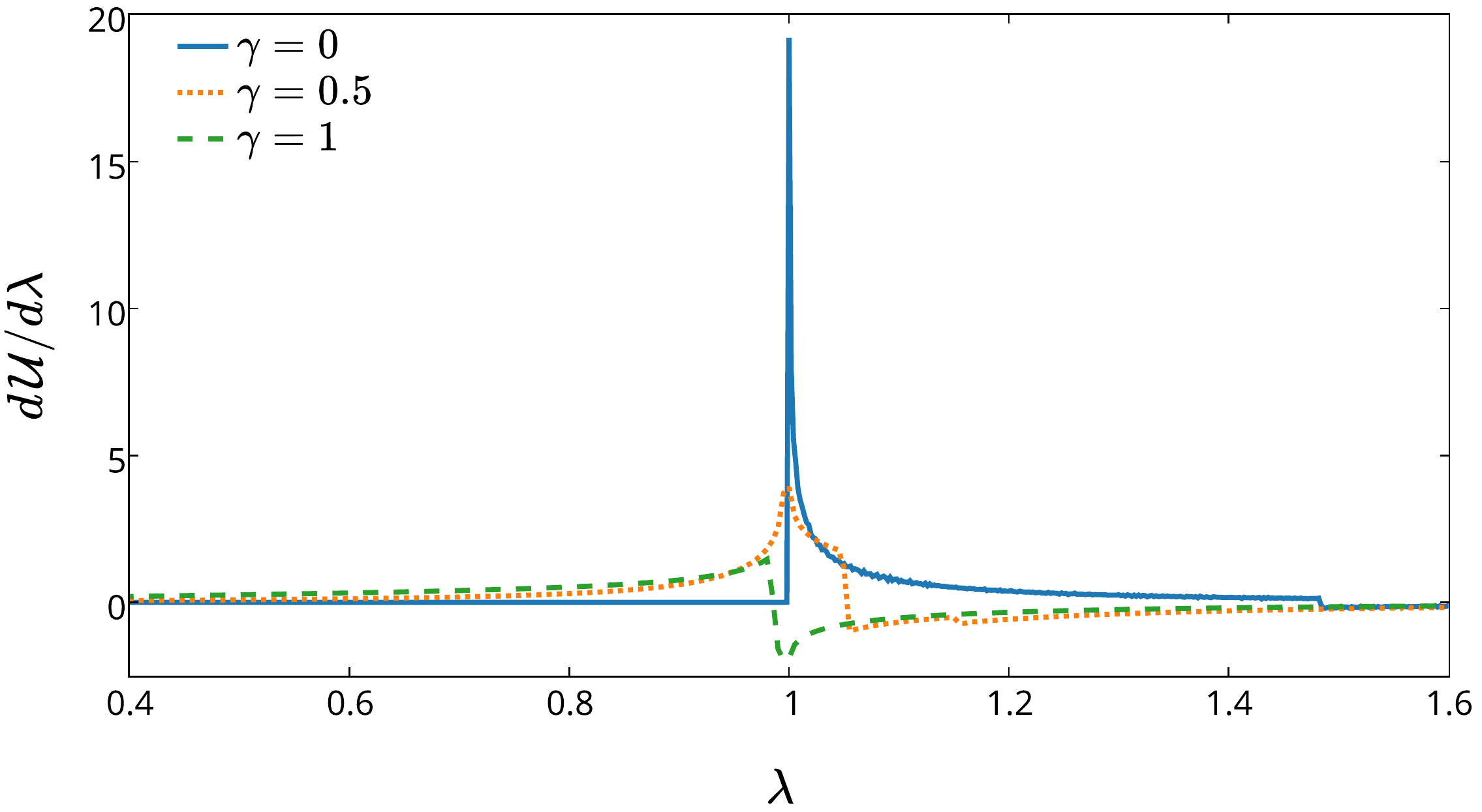}
\end{center}
\caption{Local quantum uncertainty for three distinct instances of the $XY$ model: The $XX$ model ($\gamma=0$), one example of the Ising universality class ($\gamma=0.5$) and the Ising Hamiltonian in a transverse field ($\gamma=1$). The critical point $\lambda = 1$ can be clearly identified by $\mathcal{U}$, as confirmed by its first derivative, specially for the case of the $XX$ model.}
\label{lqu_xy}
\end{figure}

Figure \ref{lqu_xy} shows the local quantum uncertainty and its derivative for three relevant cases for the $XY$ model, considering first-neighbours interactions. As we can see, the QPT at the critical point $\lambda=1$ can be clearly identified by $\mathcal{U}$, as confirmed by its first derivative. The more pronounced identification is obtained by the $XX$ model. We also observe that, beyond this divergence in the derivative of $\mathcal{U}$ at the critical point, other discontinuities points are present. These points are related with the minimization process involved in the definition of $\mathcal{U}$ and not with a QPT, which is related with a divergence presented in the derivative of the density matrix elements, as occurs at the critical point. The results presented in this section are in complete agreement with the study recently reported in Ref. \cite{fanchini}, where the local quantum coherence (as measured by the Wigner-Yanasse information) and the local quantum uncertainty were studied for the $XY$ model considering the cases $\gamma=0.5$ and $\gamma=1$.

\begin{figure}[h]
\begin{center}
\includegraphics[scale=0.37]{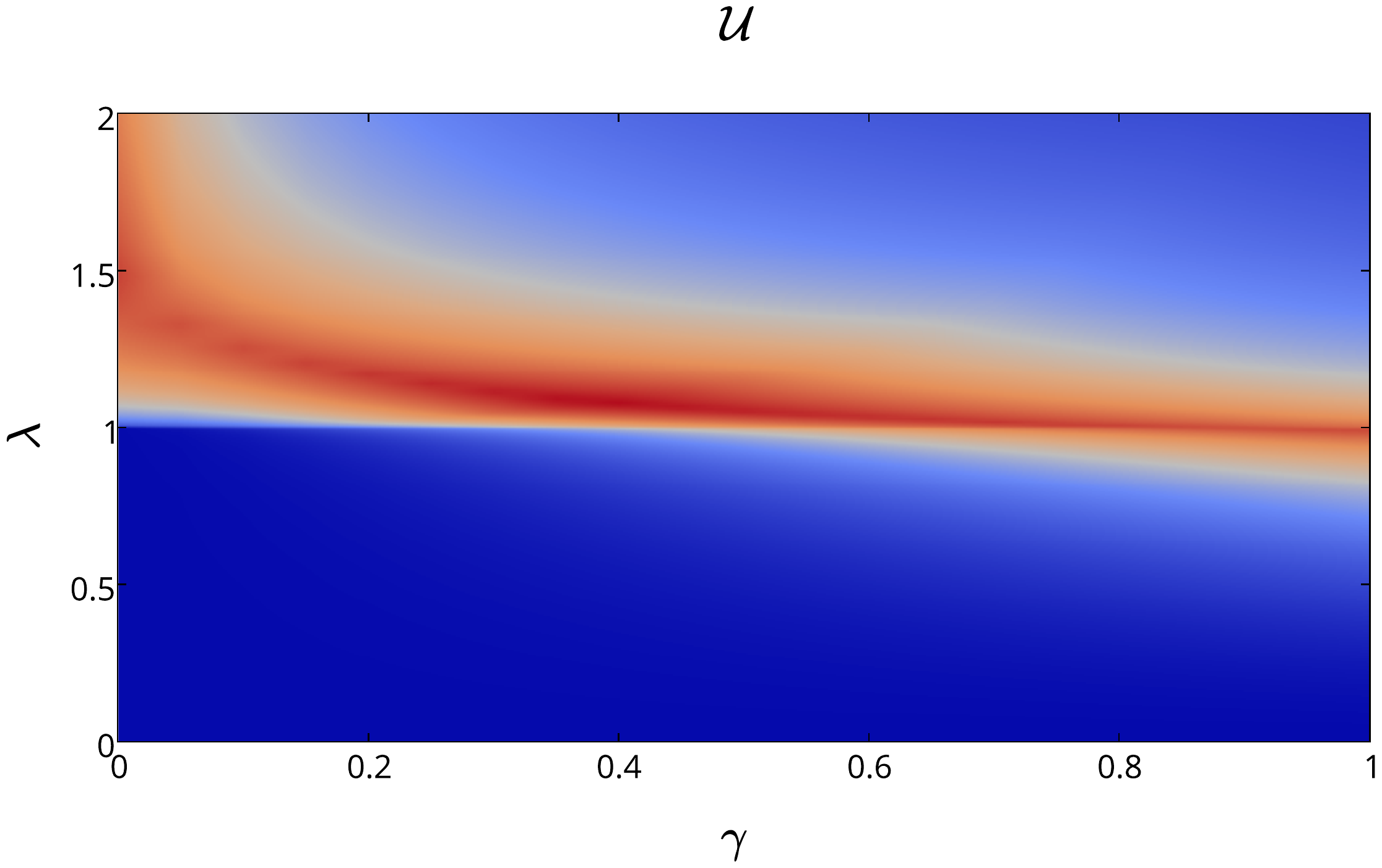} \\ \vspace{0.5cm}
\includegraphics[scale=0.37]{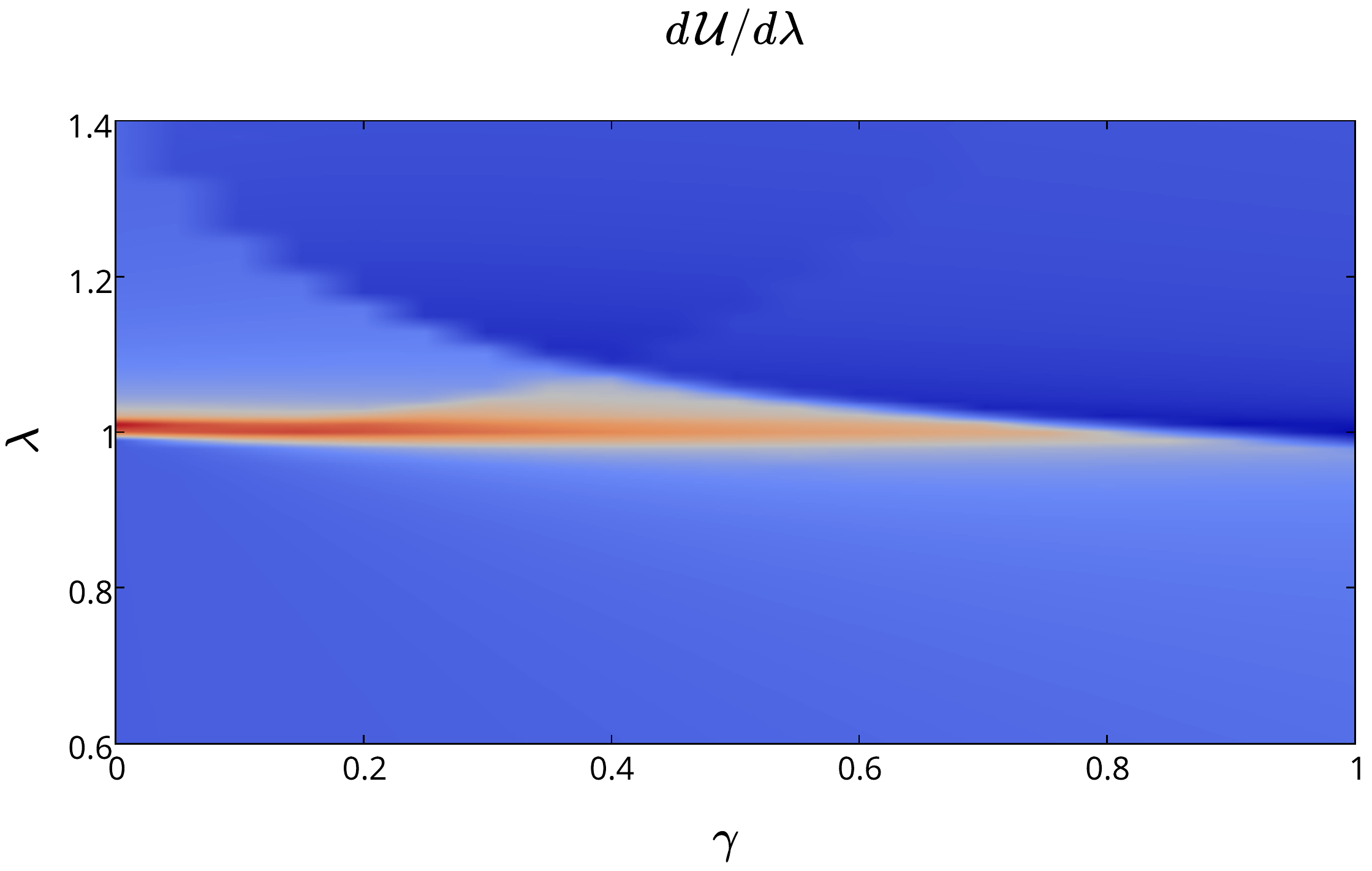}
\end{center}
\caption{Local quantum uncertainty along with its derivative as function of $\lambda$ and $\gamma$ for the $XY$ model. Although presenting different intensities, the critical line $\lambda = 1$ can be clearly identified by the derivative of $\mathcal{U}$ for all values of $\gamma$. The intensities of the quantities increase from the blue color to the red one.}
\label{lqu_xy_lg}
\end{figure}

To finish our analysis of this model, in Fig. \ref{lqu_xy_lg} we show the behaviour of $\mathcal{U}$, along with its first derivative with respect to $\lambda$, as function of $\lambda$ and $\gamma$. It is clear that the critical line $\lambda=1$ can be identified by $\mathcal{U}$ in the entire range of $\gamma$, as confirmed by its derivative.

Although the analysis presented in this section were carried out considering only first-neighbours interactions, we have also performed the same calculations considering interactions up to the 20th-neighbour interaction and the qualitative results obtained were the same, thus confirming the ability of $\mathcal{U}$ to signal the critical behaviour of this model.

\subsection{The $XY$ with three spins interaction}

As stated in the introduction, the $XYT$ model is an extended $XY$ model with includes a three-spin interaction, governed by the Hamiltonian
\begin{eqnarray}
H &=& -\frac{1}{2}\sum_{j=1}^{N}\left[(1+\gamma)\sigma^{x}_{j}\sigma^{x}_{j+1}+(1-\gamma)\sigma^{y}_{j}\sigma^{y}_{j+1} \right.\nonumber\\ 
&+&\left. \lambda\sigma^{z}_{j}+\alpha\left(\sigma^{x}_{j-1}\sigma^{x}_{j+1}+\sigma^{y}_{j-1}\sigma^{y}_{j+1}\right)\sigma^{z}_{j}\right],
\label{HXXT}
\end{eqnarray}
with $N$, $\gamma$, and $\lambda$ defined in the preceding subsection, while $\alpha$ is the intensity of the three spin interaction. In the limit $\alpha \rightarrow 0$ we obtain the usual two spin interaction $XY$ model.

Through Jordan-Wigner and Bogoliubov transformations, the Hamiltonian (\ref{HXXT}) can also be diagonalized in terms of spinless fermions operators \cite{sachdev}
\begin{equation}
H = \sum_{k=-M}^{M}2\varepsilon_{k}(d_{k}^{\dagger}d_{k}-1/2),
\end{equation} 
where $d_{k}$ satisfies the anti-commutator relations for fermions and $M=N/2$ for even $N$ or $M=(N-1)/2$ for odd $N$. 
$\varepsilon_{k}=\sqrt{\zeta_{k}^{2}+(\gamma \sin x_{k})^{2}}$, $\zeta_{k}=\lambda-\cos x_{k}-2\alpha\cos 2x_{k}$ and $x_{k}=2\pi k/N$. Considering the case of global thermal equilibrium state with inverse temperature $\beta$ (Boltzmann constant equal to one), the reduced density matrix of two arbitrary spins in the chain is given by Eq. (\ref{densityXY}), but with different formulas for the correlation functions \cite{li}. Such formulas are given in Appendix B.

\begin{figure}[h]
\begin{center}
\includegraphics[scale=0.32]{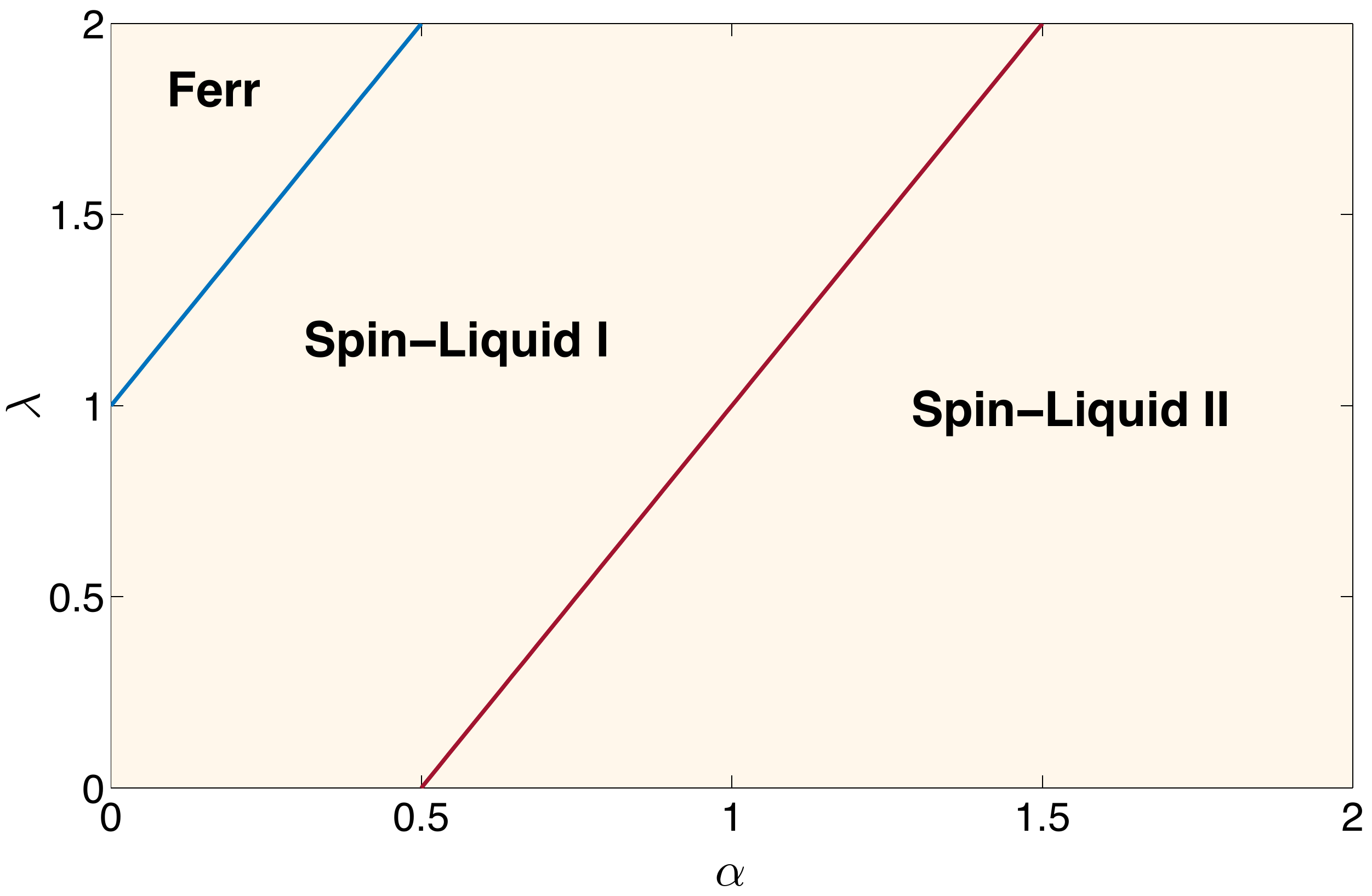}
\end{center}
\caption{Phase diagram for the $XYT$ model where the Ferromagnetic and the two spin-liquid phases are shown \cite{suzuki}.}
\label{phase_xyt}
\end{figure}

The critical properties of the $XYT$ model were studied in Ref. \cite{suzuki} and are depicted in Fig. \ref{phase_xyt} where the ferromagnetic and two spin-liquid phases can be identified. It is interesting to note that these phases are independent of $\gamma$, i.e., the phase diagram of Fig. \ref{phase_xyt} is the same independently of the value of $\gamma$. Therefore, to study the behaviour of the local quantum uncertainty near the critical points of this model, it is enough to pick just a single value of $\gamma$. Of course, the value of $\mathcal{U}$ (and of its derivative with respect to $\lambda$) will depend on $\gamma$, but the qualitative behaviour, the one which we are interested here, will be the same. In other words, if $\mathcal{U}$ is able to signal the phase transition for a single value of $\lambda$, it will be for all other values, just the intensities will change.

\begin{figure}[t]
\begin{center}
\includegraphics[scale=0.37]{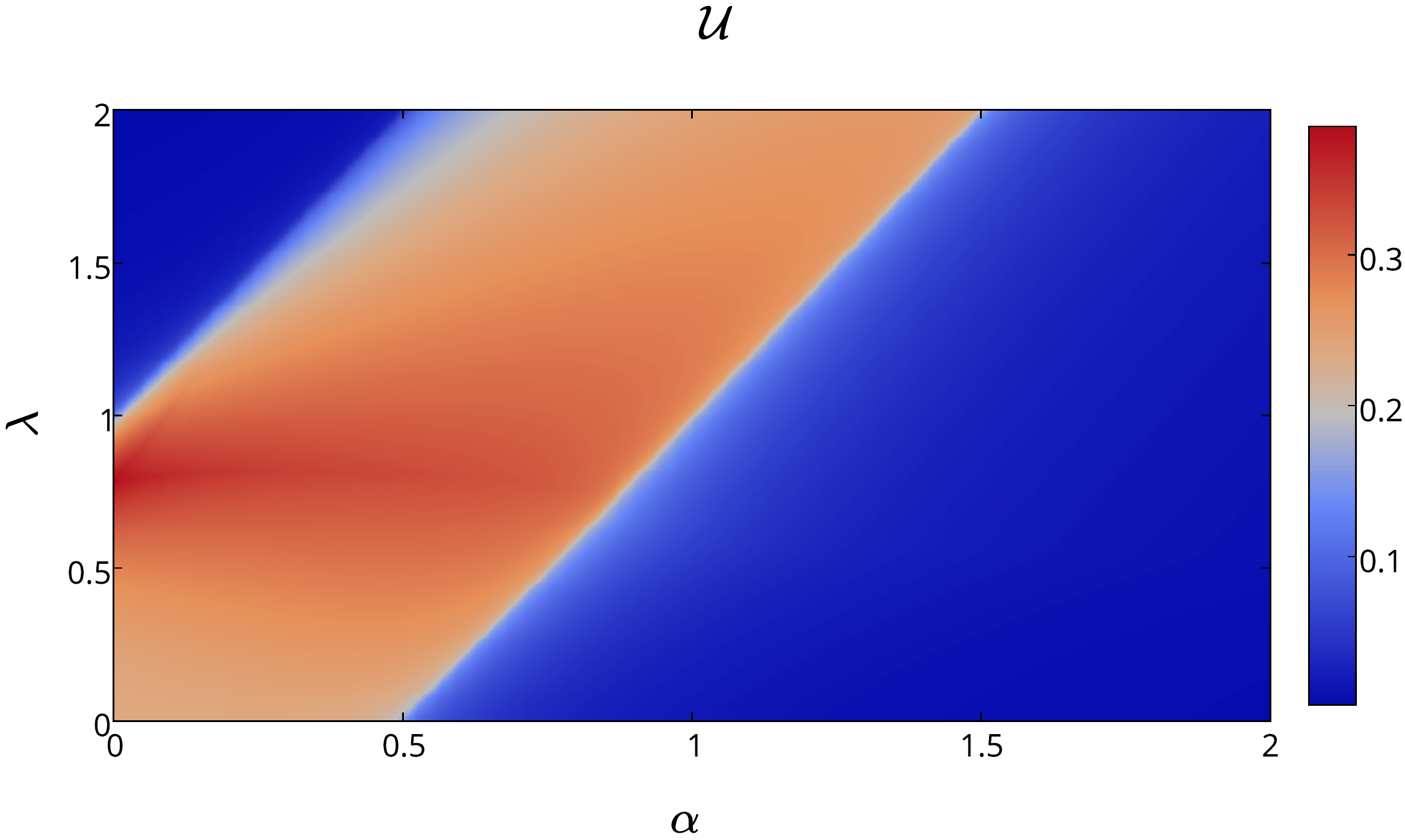} \\ \vspace{0.5cm}
\includegraphics[scale=0.37]{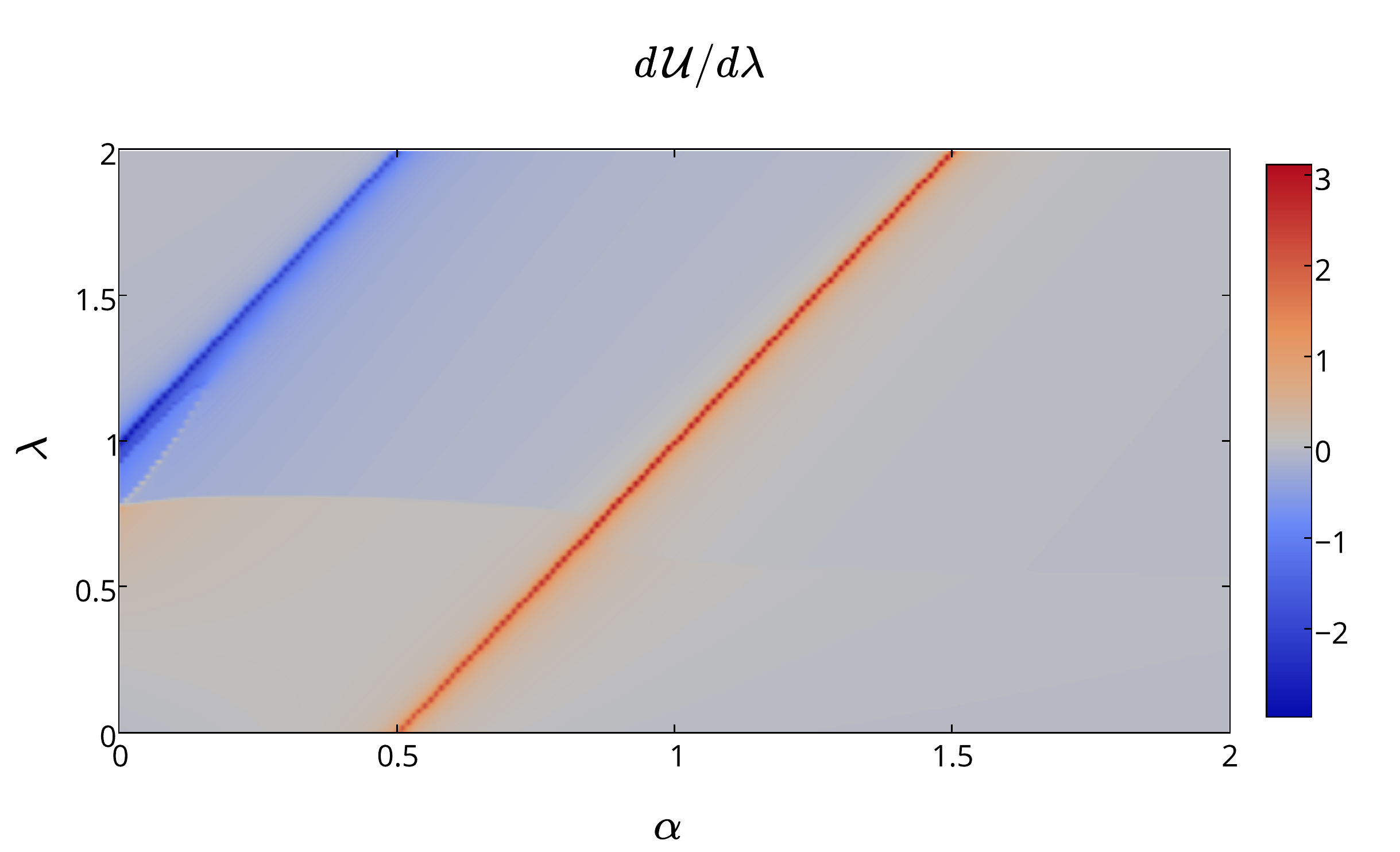}
\end{center}
\caption{Local quantum uncertainty along with its derivative as function of $\lambda$ and $\alpha$ for the $XYT$ model. We consider zero temperature, only the first-neighbour interaction, $N=2000$ and $\gamma=0.5$.}
\label{lqu_xyt}
\end{figure}

Figure \ref{lqu_xyt} shows the variation of $\mathcal{U}$ along with its derivative for the $XYT$ model. To make these figures we chose $\gamma=0.5$, $N=2000$ and considered the zero temperature case with first-neighbour interaction. As we can clearly see from the figure, the first derivative of the local quantum uncertainty is able to signal both critical lines that appear in Fig. \ref{phase_xyt}. Although not shown here, we performed several numerical studies considering different chain sizes as well as neighbour interactions other than the first. The results show that qualitative behaviour depicted in Fig. \ref{lqu_xyt} do not change, only the intensities vary, i.e., $\mathcal{U}$ is perfectly able to signal the critical lines appearing in Fig. \ref{phase_xyt} in all cases.    

\section{Discussions}

In this paper we had addressed the local quantum uncertainty in critical systems described by the $XY$ model with two and three spins interaction, the so-called $XYT$ model. Our results show that the LQU is a very useful tool in the investigation of critical behaviour. Through divergences of the derivative of $\mathcal{U}$, the critical lines of both models could be identified.

All the analysis reported here were performed considering the zero temperature case and only first-neighbours interactions. However, we also performed extensive numerical calculations for different neighbours, obtaining the same qualitative behaviour. The only change was in the intensities of the considered quantities, $\mathcal{U}$ and its derivative. While the $XY$ model was investigated in the thermodynamic limit, for the $XYT$ we considered a finite spin chain. Again, the size of the chain only changes the quantitative character of the results, with the qualitative behaviour, i.e., the ability of $\mathcal{U}$ to signal the phase transitions of the model, being the same. 

It is important to observe here that it is not always true that a sudden change in the behaviour of $\mathcal{U}$ is connected with a quantum phase transition. As a consequence of the extremization process involved in the definition of $\mathcal{U}$, the optimal observer can change while a Hamiltonian parameter is continuously modified, as also occurs with other measures of correlations \cite{sudden1,sudden2} (see also the discussion in Ref. \cite{fanchini}). We have also analysed the case of finite temperature (see \cite{werlang,maziero} for the quantum discord case). However, in this case, the divergence of the derivative (with respect to $\lambda$) of the local quantum uncertainty appears only as a local maximum at the critical point $\lambda_{c}$. We can then look at these points in order to signal the quantum phase transition.

Due to the connection of $\mathcal{U}$ with the quantum Fisher information, the results presented here indicates that critical systems could be explored in the field of quantum metrology \cite{giovannetti,marzolino}. Moreover, the connection with thermodynamics \cite{lucas} can shed fresh light on questions about the irreversible entropy production \cite{lutz} or the work distribution \cite{john} in critical systems.

\begin{acknowledgements}
It is a pleasure to acknowledge Diogo O. Soares-Pinto for critical reading the manuscript. ATA is supported by CNPq under grant numbers 459339/2014-1 and 306593/2011-4, and FAPEG under grant number 201210267001157. LCC is supported by CNPq under grants number 445516/2014-3, 401230/2014-7 and 305086/2013-8. This work was also supported by the Brazilian National Institute for Science and Technology of Quantum Information (INCT-IQ), under process number 2008/57856-6, and by CAPES.
\end{acknowledgements}

\appendix

\section{Correlations for the $XY$ model}

As stated in the main text, we consider here the $XY$ model in the thermodynamic limit ($N\rightarrow \infty$). The one-site $z-$magnetization is given by \cite{osborne}
\begin{equation*}
\langle \sigma^{z}\rangle = -\int_{0}^{\pi}\frac{(1+\lambda\cos\phi)\tanh(\beta\omega_{\phi})}{2\pi\omega_{\phi}}d\phi,
\end{equation*} 
with $\omega_{k}=\sqrt{(\gamma\lambda\sin\phi)^{2}+(1+\lambda\cos\phi)^{2}}/2$, $\beta=1/k_{B}T$ and $k_{B}$ is Boltzmann's constant. The two-site correlations functions are given by
\begin{equation*}
 \langle\sigma_{0}^{y}\sigma_{n}^{y}\rangle =\begin{vmatrix} G_{1} & G_{0} & \cdots & G_{-n+2} \\ G_{2} & G_{1} & \cdots & G_{-n+3} \\ \vdots & \vdots & \vdots & \vdots \\ G_{n} & G_{n-1} & \cdots & G_{1} \end{vmatrix},
 \end{equation*}
\begin{equation*}
\langle\sigma_{0}^{x}\sigma_{n}^{x}\rangle =\begin{vmatrix} G_{-1} & G_{-2} & \cdots & G_{-n} \\ G_{0} & G_{-1} & \cdots & G_{-n+1} \\ \vdots & \vdots & \vdots & \vdots \\ G_{n-2} & G_{n-3} & \cdots & G_{-1} \end{vmatrix},
\end{equation*}
and
\begin{equation*}
\langle\sigma_{0}^{z}\sigma_{n}^{z}\rangle = \langle \sigma^{z}\rangle^{2} - G_{n}G_{-n},
\end{equation*}
where
\begin{align}
G_{n} &\equiv \int^{\pi}_{0}\frac{\tanh(\beta\omega_{\phi})}{2\pi\omega_{\phi}}[\cos(n\phi)(1+\lambda\cos\phi) -\nonumber\\ 
 &\gamma\lambda\sin(n\phi)\sin\phi]d\phi.
\end{align}
The zero temperature limit, case considered in this paper, of these expressions is easily obtained by taking $\beta\rightarrow 0$.

\section{Correlation functions for the $XYT$ model}

The $z-$magnetization is given by \cite{li}
\begin{equation*}
 \langle \sigma^{z}\rangle = \frac{1}{N}\sum_{k}\frac{1}{\varepsilon_{k}}\left[\zeta_{k}\tanh\left(\beta\varepsilon_k\right)\right],
\end{equation*} 
where $N$ is the size of the chain and
\begin{equation*}
\varepsilon_{k} = \sqrt{\zeta_{k}^{2} + \gamma^{2}\sin^{2}\left(x_{k}\right)},
\end{equation*}
$\zeta_{k} = \lambda - \cos\left(x_{k}\right) - 2\alpha\cos\left(2x_{k}\right)$ and $x_{k} = 2\pi k/N$. The two-site correlation functions are defined as
\begin{equation*}
\langle\sigma_{0}^{x}\sigma_{n}^{x}\rangle =\begin{vmatrix} g_{-1} & g_{-2} & \cdots & g_{-n} \\ g_{0} & g_{-1} & \cdots & g_{-n+1} \\ \vdots & \vdots & \vdots & \vdots \\ g_{n-2} & g_{n-3} & \cdots & g_{-1} \end{vmatrix},
\end{equation*} 
\begin{equation*}
\langle\sigma_{0}^{y}\sigma_{n}^{y}\rangle =\begin{vmatrix} g_{1} & g_{0} & \cdots & g_{-n+2} \\ g_{2} & g_{1} & \cdots & g_{-n+3} \\ \vdots & \vdots & \vdots & \vdots \\ g_{n} & g_{n-1} & \cdots & g_{1} \end{vmatrix},
\end{equation*}
and
\begin{equation*}
\langle\sigma_{0}^{z}\sigma_{n}^{z}\rangle = \langle \sigma^{z}\rangle^{2} - g_{n}g_{-n},
\end{equation*}
with
\begin{align*}
g_{n}&=-\sum_{k}\frac{1}{N\varepsilon_{k}}\left[\cos(x_{k}n)\zeta_{k}+\gamma\sin(x_{k}n)\sin(x_{k})\right]\times \nonumber\\ 
&\tanh\left(\beta\varepsilon_k\right).
\end{align*}

Note that the above expressions, for the limit $\alpha\rightarrow 0$ give us the relevant correlation functions for the $XY$ model without taking the thermodynamics limit. As in the preceding case, the zero temperature limit is easily obtained here. 


\end{document}